# Compressive dual-comb spectroscopy

A<small>KIRA</small> K<small>AWAI</small>[1], T<small>AKAHIRO</small> K<small>AGEYAMA</small>[1], R<small>YOICHI</small> H<small>ORISAKI</small>[2,3], <small>AND</small> T<small>AKURO</small> I<small>DEGUCHI</small>[1,3,4]

[1]*Department of Physics, The University of Tokyo, Tokyo, Japan*

[2]*Graduate School of Information Science and Technology, The University of Tokyo, Tokyo, Japan*

[3]*PRESTO, Japan Science and Technology Agency, Saitama, Japan*

[4]*Institute for Photon Science and Technology, The University of Tokyo, Tokyo, Japan*

*ideguchi@ipst.s.u-tokyo.ac.jp

**ABSTRACT**

**Broadband, high resolution and rapid measurement of dual-comb spectroscopy (DCS) generates a large amount of data stream. We numerically demonstrate significant data compression of DCS spectra by using a compressive sensing technique. Our numerical simulation shows a compression rate of more than 100 with 3% error in mole fraction estimation of mid-infrared (MIR) DCS of two molecular species in a broadband (~30 THz) and high resolution (~115 MHz) condition. We also numerically demonstrate a massively parallel MIR DCS spectrum of 10 different molecular species can be reconstructed with a compression rate of 10.5 with a transmittance error of 0.003 from the original spectrum.**

## 1. Introduction

In the last decade, intensive attention has been cast on dual-comb spectroscopy (DCS), which allows us to measure broadband and high resolution spectra with superior frequency accuracy at a high data acquisition rate [1,2]. DCS provides unprecedented spectroscopic features especially for multiplex gas-phase molecular sensing with its sub-Doppler spectral resolution of ~100 MHz spanning over ~10s THz, enabling various applications such as precision metrology [3], greenhouse gas sensing [4,5], combustion diagnosis [6] etc. The broadband and high-resolution spectroscopy can generate a large data set, e.g., 1,000,000 spectral points for a single spectrum [7]. Now, if we imagine the DCS techniques are to be used for hyperspectral imaging of 1,000 × 1,000 pixels measured with a 16 bit analog-to-digital converter, it generates ~4 TB per single hyperspectral image. Taking such images would cause severe problems in data transportation and/or storage.

Compressive sensing (CS) is a signal processing technique that allows, by making use of sparsity of a signal, reconstruction of the signal from significantly reduced number of data points than the full set of data points required from the Nyquist-Shannon sampling theorem [8]. If the signal is sparse in a certain basis, the sparsest solution can be found by algorithms with a sparsity constrain or regularization. Mathematical studies proved that, in some appropriate conditions, CS can reconstruct an exact signal even in the presence of measurement noise [8]. A variety of studies on CS have been reported especially in the field of optical imaging, where natural scenes such as landscapes or biological cells are well reconstructed from images with less pixels [9-11]. Contrary to imaging, CS-based spectroscopy [12-16], especially for gas-phase molecular sensing, has not actively been investigated, although broadband high-resolution spectra of gaseous molecules are good candidates of CS because of their sparse nature due to the narrow molecular lines spread in a broad spectral range.

In this study, we numerically demonstrate compressive dual-comb spectroscopy (C-DCS), in which a simple CS technique effectively compresses the data size of DCS. In our numerical demonstration, we show that quantitative estimation of mole fraction of molecules can be made with an error of 3% even when <1% of interferometric data points are used only. In addition, we show a well reconstructed complex spectrum of mixture of 10 molecular species in a condition of using 10% of original data points.

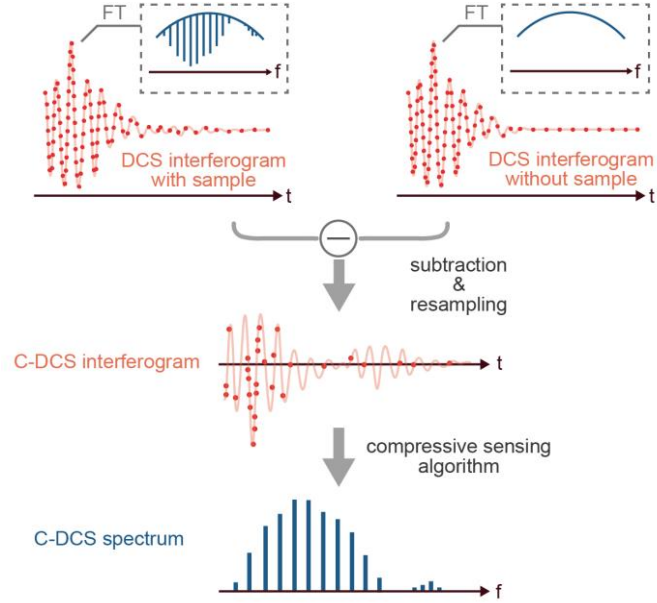

**Fig.1 A conceptual illustration of compressive dual-comb spectroscopy (C-DCS).**

## 2. Concept of compressive dual-comb spectroscopy (C-DCS)

The basic concept of C-DCS is described in Fig.1. DCS is a type of Fourier-transform spectroscopy with two mutually coherent frequency combs that run at slightly detuned repetition rates [1,2]. The spatially combined pulse trains are interferometrically detected with a single photodetector and digitization of the signal generates an interferogram. Finally, Fourier-transform of the interferogram shows a spectrum. Our simulation is conducted under an assumption that we have a full original data set of a DCS interferogram and randomly resample the data points at a lower rate (reducing the data points) to reconstruct a CS spectrum. Suppose that the number of data points of the original interferogram is $N$ and that of the resampled data points is $M$ ($M < N$), the resampled data set can be represented as $\Omega = \{\omega_j\}_{j=1}^{M} \subset \{1,2,\ldots,N\}$. Note that we can arbitrarily set a probability mass function (PMF) for the random resampling. From the resampled data set, we can estimate a spectrum vector $x \in \mathbb{C}^N$ as an answer of the $l1$ minimization problem [8] described as

$$\min\|\tilde{x}\|_1 \quad \text{subject to} \quad \|A\tilde{x} - y\|_2 < \epsilon \quad (1),$$

where $A \in \mathbb{C}^{M \times N}$ ($M < N$) is a sensing matrix, $y \in \mathbb{R}^M$ a measurement vector, and $\epsilon \in \mathbb{R}$ a constraint value determined by noise of the system. In our case, $A = R\Phi$, where $R \in \{0,1\}^{M \times N}$ is a subsampling operator which obeys $(Rx)_j = x_{\omega_j}$, and $\Phi$ a $N \times N$ discrete Fourier-transform operator. Considering a case of DCS operated with relatively low-chirped combs, signal intensities of the sampled points around the zero delay between the pulses have a larger magnitude (called

"center-burst") than the other points including the signals showing molecular induction decays [17]. Therefore, it is effective to select a sloped PMF that samples more points around the center-burst.

To fully utilize the sparse nature of absorption spectrum and efficiently compress the data points, we suggest to operate background subtraction of the interferogram. It can be implemented either by a hardware instrumentation or a post numerical processing. For the hardware instrumentation, a Michelson-type interferometer is added so as to make the $\pi$ phase difference between the pulses from the two arms due to the reflection of the beam splitter, realizing the background subtraction due to the destructive interference on the detector [18]. On the other hand, for the post numerical processing of background subtraction, a reference background interferogram can be obtained either by an additional measurement [19] or a numerical baseline reconstruction [4]. Although in this proof-of-concept demonstration we operate the background subtraction for better reconstruction, we expect improved algorithms would make possible to reconstruct spectra with background.

## 3. Numerical condition of C-DCS simulation

To show the above-mentioned C-DCS concept, we demonstrate numerical simulations of trace-gas DCS in the MIR region. We simulate a mimic condition of a previously reported experiment [19], where a broadband spectrum covering from 2006.7 to 3013.4 cm$^{-1}$ (60.159-90.339 THz) is measured at a resolution of 0.0038 cm$^{-1}$ (115 MHz) that consists of 262,144 spectral points Fourier-transformed by temporal data points of 524,286. We assume a broadband Gaussian-profile spectrum as comb sources. We first simulate an interferogram from the source spectrum with molecular absorptions with Doppler line profiles and create a background-free interferogram by baseline subtraction with a reference spectrum with no absorption lines. We numerically calculate the spectrum by referencing HITRAN database and using its application programming interface HAPI [20]. Then, the interferogram is resampled with a sloped PMF, $C \min\{1, 1/|l - N/2|\}$, where $l$ is an index of sampling points of the intereferogram in chronological order and $C$ a normalization constant, which is found in literature [21]. For spectrum reconstruction, we use SPGL1 as a $l1$- minimization problem solver [22]. We set an arbitral signal-to-noise ratio (SNR) by assuming coherently averaged interferograms, which can be experimentally implemented with a variety of techniques [7,23,24]. We note that the CS reconstruction algorithm in general works without having pre-knowledge of the molecular species.

## 4. Mole fraction estimation of two molecular species

To show how the data compression rate, which is defined as $N/M$, affects quantitative capability of C-DCS, we demonstrate mole fraction estimation of two molecular species of trace gases. We numerically prepare mixed gases of N$_2$O (42 ppm) and CO (120 ppm) with a buffer gas at a pressure of 3 mbar filled in a 10-m-long multi-pass cell. We add a Gaussian measurement noise $\boldsymbol{n}$ to the interferogram so as to make the estimation with different SNR conditions. We first set SNR of the real part of FFT to be 1,000. A constraint term $\epsilon$ in Equation (1) is empirically set to an average of $||\boldsymbol{n}/10||$. The original spectrum converted from the full data points of 524,286 and compressive spectra with 10,000 and 2,000 sampling points are shown in Fig. 2, where we show transmittance spectra of the sample. The compressive spectrum with 10,000 sampling points, which compression rate is 52.4, shows good agreement with the original one, while that with 2,000 (compression rate of 262) shows clear distortions. We evaluate mole fraction of N$_2$O molecules by spectrally fitting each absorption line and obeying Lambert-Beer law. The fitting is operated by a fixed-profile Gaussian function with a single free parameter of mole fraction. Here, the spectral points that satisfy $\log(1 - T) > 0.01$ (T: transmittance) are used for the evaluation. Figure 3 (a) shows ratio of the evaluated mole fraction of the compressed and original spectra for different compression rates under different SNR (1,000, 500, 100) conditions. The result with SNR of 1,000 shows that the compression rate of 105, which corresponds to the number of sampling point of 5,000, leads to 3% of deviation of mole fraction from that evaluated with the original spectrum. We can also see that lower measurement SNR degrades the evaluation results. Figure 3 (b) shows root-mean-squared error (RMSE) of the spectral points that satisfies $\log(1 - T) > 0.01$ for each compression rate. We find that RMSE is proportional to (compression rate)$^{0.93}$ by least squares fitting.

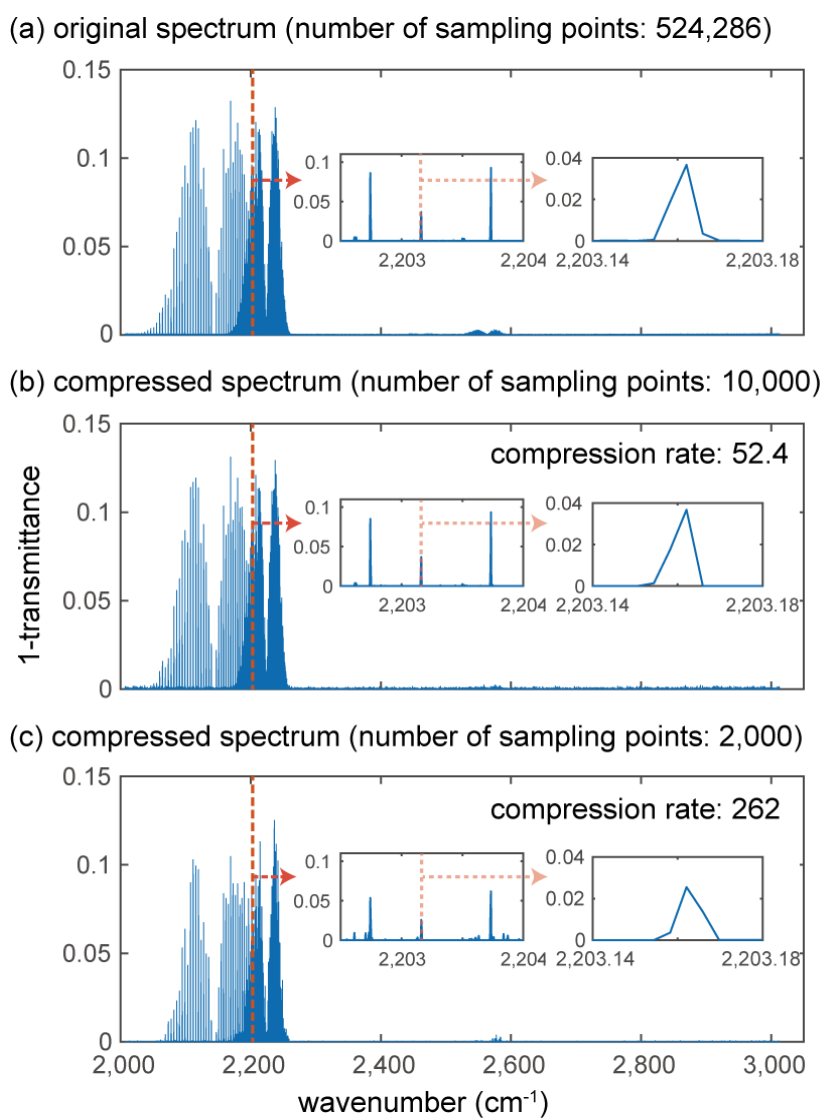

Fig.2 Simulated MIR DCS spectra of gaseous molecules. (a) An original spectrum Fourier-transformed by temporal data points of 524,286. (b), (c) Compressed spectra with 10,000 and 2,000 sampling points, respectively. The inset panels show zoom-in spectra around 2,203 cm$^{-1}$.

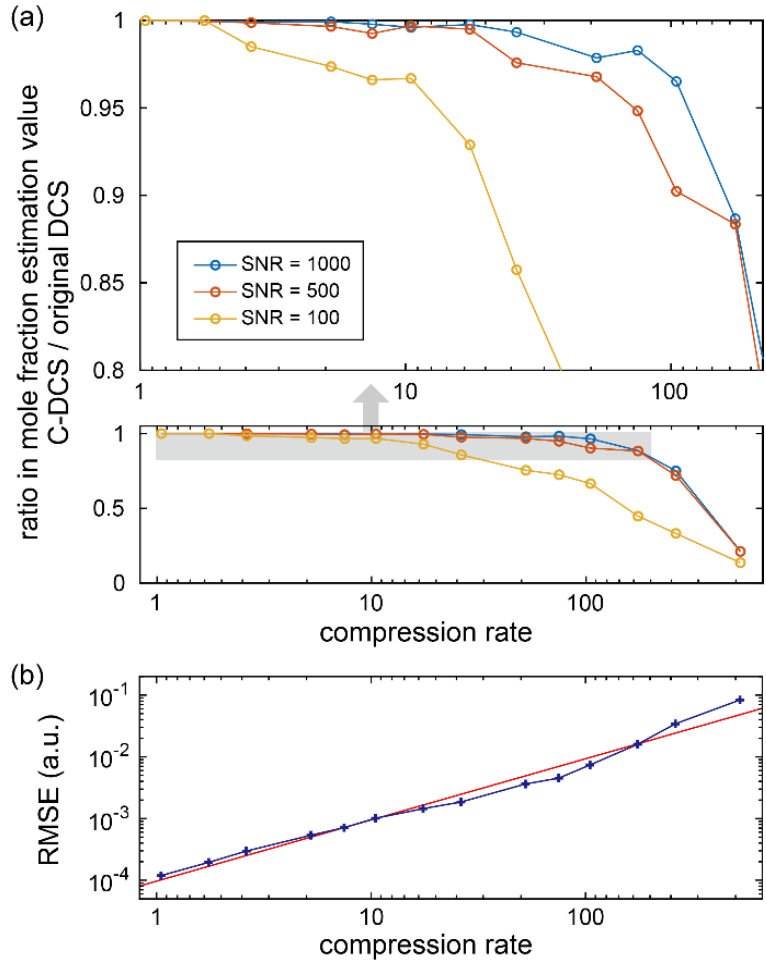

**Fig.3 (a) Ratio in mole fraction estimation value of C-DCS and original DCS with different compression rates. Blue, Red, and Yellow data show the results with different SNR. (b) Root-mean-squared error (RMSE) of the spectral points for each compression rate. The fitting (red line) shows that RMSE is proportional to (compression rate)$^{0.93}$.**

## 5. Robustness evaluation of C-DCS reconstruction

We quantitatively evaluate the robustness (deviation from the ground truth) of the CS reconstruction in terms of the peak transmittance, center frequency and linewidth of an absorption line by simulating spectra with 100 different patterns of random sampling for each condition. We analyze a single absorption line of $N_2O$ at 2238.36 cm$^{-1}$ in the spectra calculated in the same condition as that shown in Fig. 2 with the SNR of 1,000. Here we change the sample lengths (76, 11.4, 1.67, 0.76, 0.15 m) so that we can see how the absorption peak transmittance affects the CS reconstruction quality. Figure 4(a) shows the peak transmittance of the absorption line as a function of the compression rate. The standard deviations of the 100 spectra calculated with the different random saplings are illustrated as weak color bands around the mean values. Figure 4(b) shows the mean value divided by the standard deviation of the data points shown in Fig. 4(a). It clearly shows the reconstruction degrades at higher compression rates and the absorption lines with higher absorption intensities (longer sample lengths) are reconstructed more robustly. Figure 4(c) shows the deviation in center wavenumber of the absorption line from the ground truth. The mean values are mostly within the spectral resolution (0.0038 cm$^{-1}$), showing its high robustness in the CS reconstruction. Figure 4(d) shows the relative linewidth to the ground truth. Although the absorption lines with higher peak intensities are well reconstructed up to the compression rate of ~100, the ones with lower peak intensities are largely deviated from the ground truth at the higher compression rates.

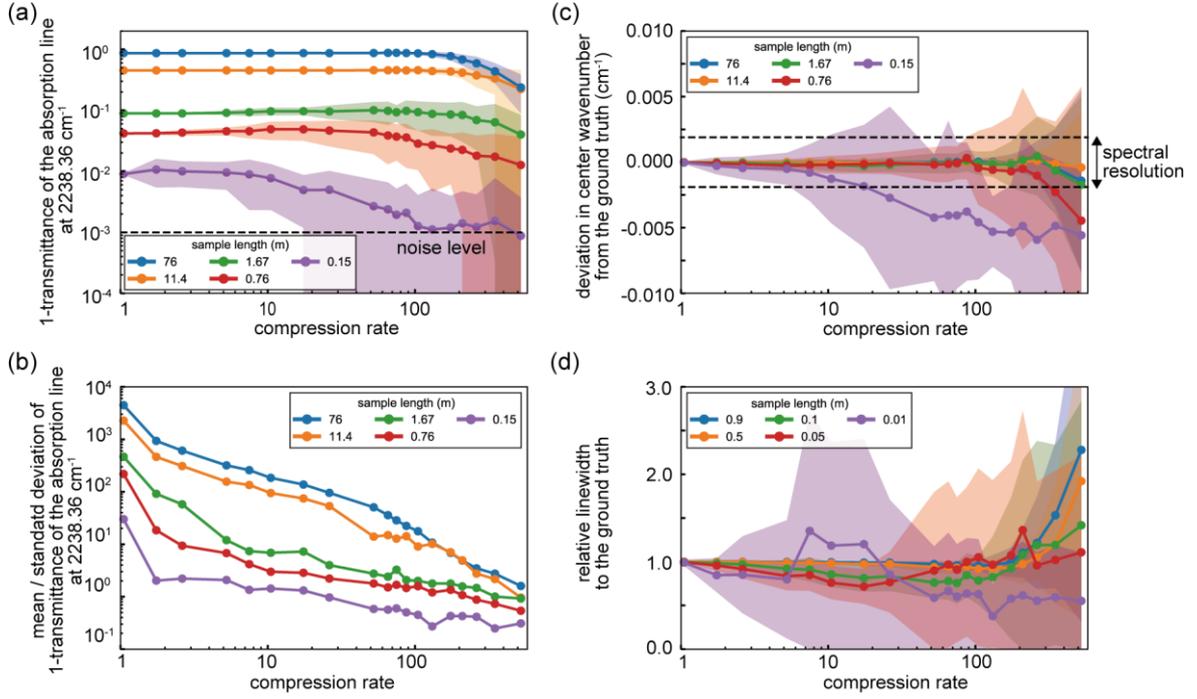

**Fig.4 Robustness evaluation of CS reconstruction as a function of the compression rate in terms of (a) peak transmittance, (b) mean / standard deviation of (a), (c) deviation in center wavenumber from the ground truth, (d) relative linewidth to the ground truth, of the absorption line at 2238.36 cm$^{-1}$. The plots with different colors show the calculation results with different sample lengths (blue: 76 m, yellow: 11.4 m, green: 1.67 m, red: 0.76 m, purple: 0.15 m). The dots and bands represent the mean values and standard deviations of the 100 spectra calculated with the different random samplings.**

## 6. Massively parallel C-DCS of 10 trace gas species

Finally, to show the compression capability of C-DCS for denser molecular lines, we demonstrate massively parallel spectroscopy of 10 trace gas species, which resembles to the previously reported experiment [18]. We assume a 76-m-long multi-pass gas cell filled with nitrous oxide ($^{14}N_2^{16}O$) at 42 ppm, nitric oxide ($^{14}N^{16}O$) at 420 ppm, carbon monoxide ($^{12}C^{16}O$) at 120 ppm, carbonyl sulfide ($^{16}O^{12}C^{32}S$) at 26 ppm, methane ($^{12}CH_4$) at 1,500 ppm, ethane ($^{12}C_2H_6$) at 490 ppm, ethylene ($^{12}C_2H_4$) at 540ppm, acetylene ($^{12}C_2H_2$) at 6,600 ppm, carbon dioxide ($^{12}CO_2$, $^{13}CO_2$), at 280 ppm, water vapor ($H_2^{16}O$) at 2,100 ppm and a buffer gas. The most abundant isotope of each element except for $CO_2$ is included in this simulation. The pressure is set to 3 mbar. To relax regularization and let small absorption peaks remain, we increase the constraint term $\epsilon$ to the average of $||\boldsymbol{n}||$. Figure 5 shows the C-DCS spectra with a compression rate of 10.5. We clearly observe that vibrational absorption lines are well reconstructed compared to the original ones with the error (standard deviation) in transmittance of less than 0.003.

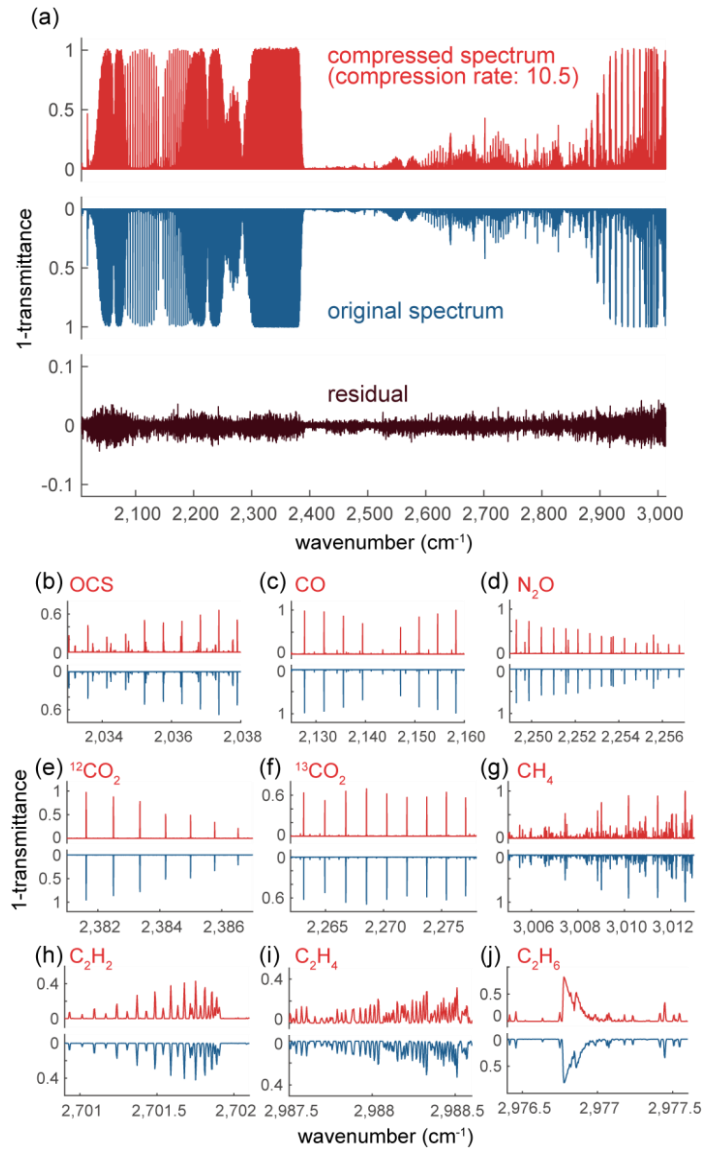

**Fig.5 Massively parallel spectra of 10 trace gas species.** (a) Comparison between a compressed spectrum and an original spectrum. The compressed spectrum is reconstructed from 50,000 points (compression rate of 10.5). (b-j) Zoom-in spectra of characteristic absorption lines of different molecular species.

## 7. Discussion

The concept of C-DCS can be applicable to other Fourier-transform spectroscopy (FTS) including Michelson-type FTS, FT-Raman spectroscopy [25], and FT-CARS spectroscopy [26,27] etc. It can also be used for a spectrum measured in the spectral domain by calculating an interferogram by Fourier-transforming the spectrum. The performance of C-DCS can be improved by using other PMFs and/or reconstruction algorithms especially developed for the use of CS imaging. The compression of C-DCS would become more valuable when we use it for higher dimensional measurements such as hyperspectral imaging [28] or multi-dimensional DCS [29]. Lastly, we note that C-DCS is possibly to be used for speeding up DCS measurement by implementing the compressive sampling in hardware. For that purpose, for example, we can arbitrary sweep the difference in repetition rate during a measurement of an interferogram, allowing a non-uniform temporal waveform sampling.

**Funding.** Precursory Research for Embryonic Science and Technology (JPMJPR17G2); Japan Society for the Promotion of Science (20H00125).

**Acknowledgments.** R. H. and T. I. thank Japan Science and Technology Agency for their support.

**Disclosures.** The authors declare no conflicts of interest.